\documentclass[letterpaper,twocolumn,10pt]{article}
\usepackage{usenix,epsfig}

\usepackage{gensymb}
\usepackage{xspace}
\usepackage{caption}
\usepackage{subcaption}
\usepackage{enumerate}
\usepackage{enumitem}
\usepackage{cite}

\usepackage[hyphens]{url}
\usepackage[hidelinks]{hyperref}
\hypersetup{breaklinks=true}

\usepackage{amsmath}
\usepackage{algorithm}
\usepackage[noend]{algpseudocode}
\usepackage{float}

\usepackage{listings}
\usepackage{color}

\definecolor{dkgreen}{rgb}{0,0.6,0}
\definecolor{gray}{rgb}{0.5,0.5,0.5}
\definecolor{mauve}{rgb}{0.58,0,0.82}

\makeatletter
\newcounter{phase}[algorithm]
\newlength{\phaserulewidth}

\makeatother

\newcommand{\edp}{\emph{eDisco}\xspace}

\begin{document}

\date{}

\title{\Large \bf  eDisco: Discovering Edge Nodes Along the Path}

\author{
{\rm Aleksandr Zavodovski}\\
University of Helsinki
\and
{\rm Nitinder Mohan}\\
University of Helsinki
\and
{\rm Jussi Kangasharju}\\
University of Helsinki
} 

\maketitle

\thispagestyle{empty}

\subsection*{Abstract}

Edge computing is seen as an enabler for upcoming applications
requiring low latency offloading, such as augmented reality, and as a
key building block for Internet of Things.
Edge computing extends the centralized cloud computing model by
distributing servers also close to the users, at the edge of the
network.
A key challenge for the clients remains on how to discover the
nearby edge servers and how to access them.
In this paper, we present \edp, DNS-based edge discovery,
that leverages existing protocols and requires no modifications to deployed infrastructure.
\edp enables service providers and clients to discover edge servers,
and determine the optimal edge deployment configuration. 

%
%
%

\section{Introduction}
\label{sec:introduction}

The introduction of cloudlets~\cite{satyanarayanan2009case} and
fog~\cite{bonomi2012fog} served as a starting point for the recent
developments in edge computing, pushing it as a key enabler for IoT, augmented reality, vehicular networking, and numerous other more specific use cases~\cite{nokia_mec, shopping_edge}. 
%
%
By bringing the edge servers close to the users, novel applications and services requiring very low latencies become feasible, and the growing popularity of these kinds of applications is driving the development of edge computing. 
To fully leverage  edge computing, two key
challenges need addressing.

First is the placement of the edge servers which has
already seen a fair amount of research,
e.g.,~\cite{Ceselli2017MobileEC, mohan2016edge, Bahreini2017EfficientPO}. The main research questions on this topic revolve around various placement algorithms, determining good or optimal locations, and trying to approach the problem from the point of view of the edge service provider. While important to address and understand, they only cover part of the whole problem space of edge computing.

The second challenge relates to how the users and devices can
discover the edge servers, to actually use them. This is the challenge we address in this paper, and we
have identified the following requirements for an efficient discovery
solution. First and foremost, it should work out-of-the-box with
existing infrastructure, servers, and clients, in order to facilitate
the adoption of edge computing. Second, as end-to-end encryption
becomes prevalent on the Web, solutions clinging to on-path
interception of requests are not feasible, mandating an
end-to-end approach. Third, the solution should be efficient and
flexible, and allow for different applications and services to be
developed on it. In essence, we need a standardized edge discovery
protocol that matches these requirements.

%
%
%

In this paper we present \edp (Edge Discovery Protocol), a DNS-based
solution that successfully addresses the key requirements from
above. We use DNS SRV-records to indicate the locations of edge
servers, which allows both cloud and other service providers to
determine possible locations for onloading and for clients and any
other devices to locate possible offloading locations. \edp does not
require any changes to any existing infrastructure, clients, servers,
or protocols; only the entity installing an edge server needs to add
the corresponding SRV records into their DNS zone. \edp is a
light-weight and efficient solution and meets the requirements
specified above.
%
%
%
We believe that \edp can serve as an enabler for edge computing by
allowing all players to start offering edge computing services and
give a boost to edge-aware distributed application development. Further, \edp is an another step towards "overthrowing the Internet feudalism", as Liu et al. \cite{liu2017barriers} pessimistically characterize the present state of the Internet.

The rest of this paper is organized as follows. Section~\ref{sec:edp}
describes the basic design and operation of
\edp. Section~\ref{sec:edp-practicalities} discusses the practical
implementation issues of \edp, in particular in terms of collecting
the required information. In Section~\ref{sec:discussion} we discuss
\edp's applicability to various use cases and review related work in
Section~\ref{sec:related}. Finally, Section~\ref{sec:conclusion}
concludes the paper.


\section{Edge Discovery Protocol (\edp)}\label{sec:edp}

%
%
%

\begin{figure*}
	\centering
	\begin{subfigure}{.4\textwidth}
		\centering
		\includegraphics[width=.75\linewidth]{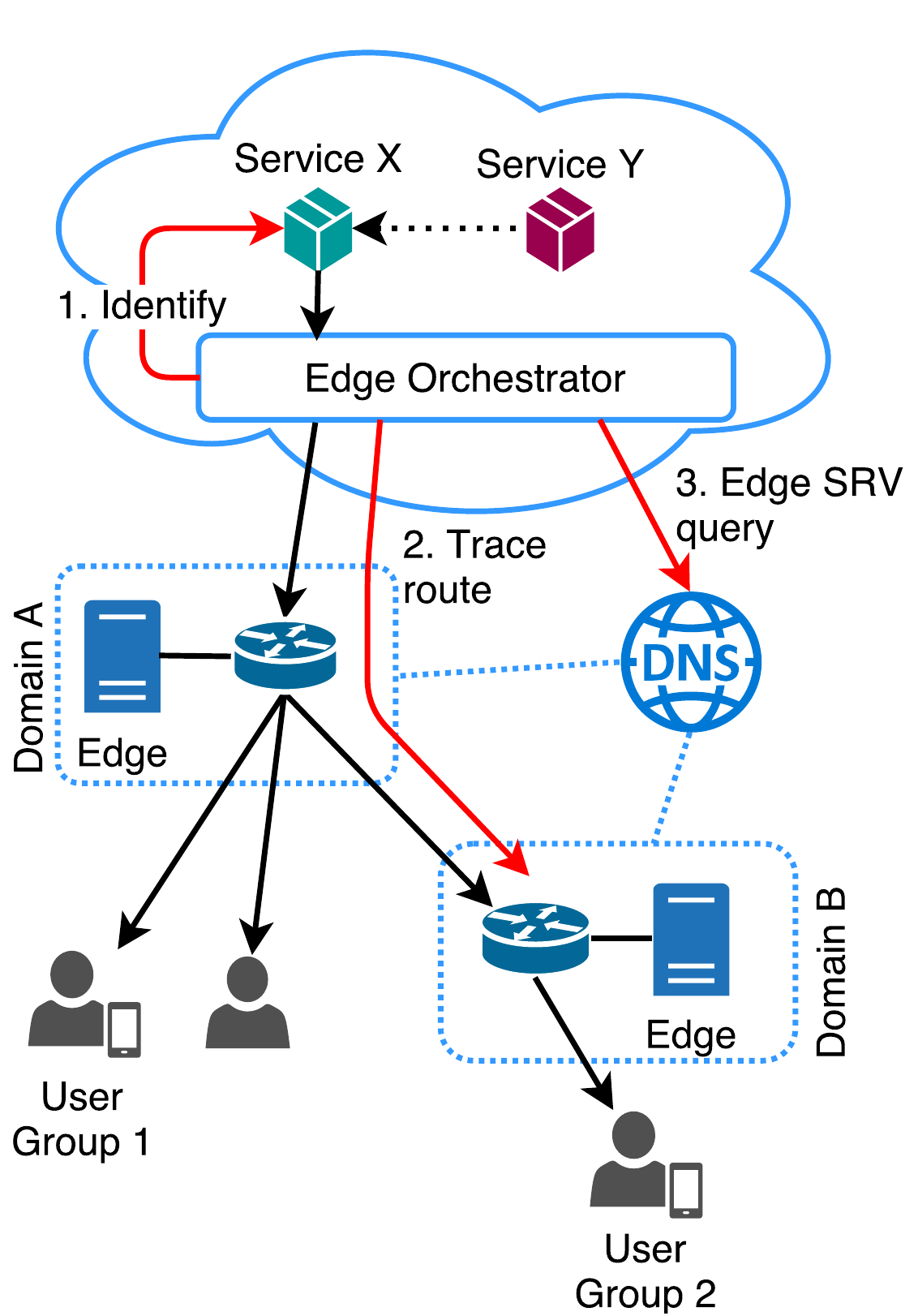}
		\caption{Edge discovery first phase, actual edge discovery.}
		\label{fig:sys-1-stage-1}
	\end{subfigure}%
	\hfill%
	\begin{subfigure}{.4\textwidth}
		\centering
		\includegraphics[width=.75\linewidth]{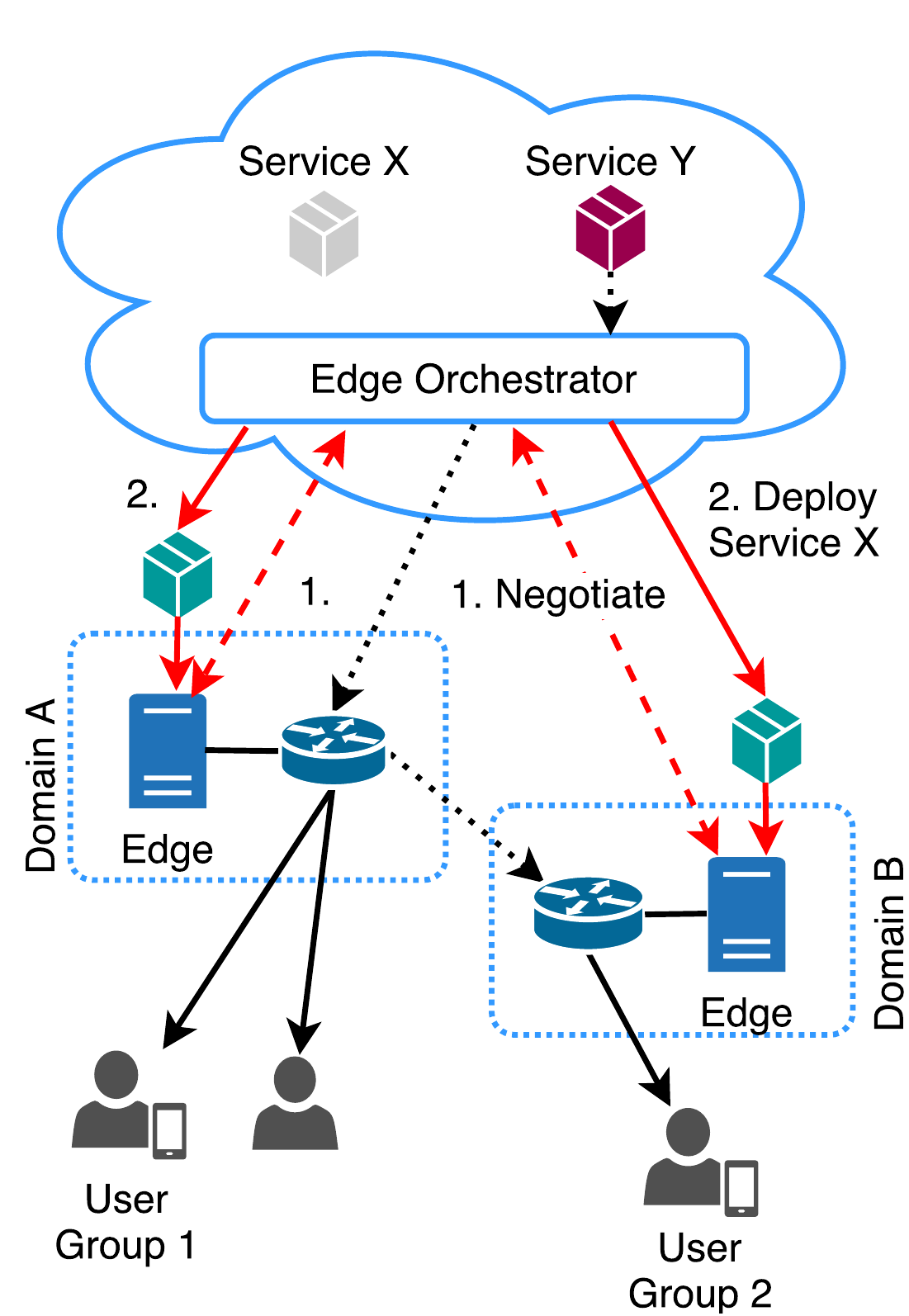}
		\caption{Edge discovery second phase, negotiation and deployment. }
		\label{fig:sys-1-stage-2}
	\end{subfigure}
	\caption{Edge discovery, onloading case. \edp actions of edge orchestrator are shown in red. Solid line is used for bandwidth-intensive traffic, dotted line for low bandwidth.}
	\label{fig:sys-1}
\end{figure*}

We introduce \edp via a concrete example of computational onloading, where service residing in a cloud is pushed closer to the end users.
We use this scenario as an example, as it illustrates all the facets
of \edp; other use cases, e.g., traditional offloading and others are
simpler and are discussed in Section~\ref{sec:discussion}.
For convenience, we split the scenario in two phases: actual discovery (Figure~\ref{fig:sys-1-stage-1}) and deployment of services
(Figure~\ref{fig:sys-1-stage-2}). 

The design goal of \edp is to allow for the service provider to
spot (all) possible locations of edge servers between its own
servers and the clients, and use this information to determine which
of these possible locations it wishes to use for deploying its
services. 

\noindent
The \textbf{first phase}, actual discovery, is the core functionality
of \edp and requires only usage of common networking protocols and
tools, namely, DNS and \texttt{traceroute}. The process is illustrated in
Figure~\ref{fig:sys-1-stage-1} and is described below in detail.
The service provider (e.g., a cloud provider) observes requests from
clients coming from different parts of the network. We assume the
provider runs an orchestrator to manage the process.
%

%

\noindent
The \textbf{second phase}, deployment, during which the service is
relocated to newly detected edge follows the discovery phase. This
part optimizes the deployment of services and handles redirection of
clients to the onloaded edge services.

While the actual client redirection happens in real-time, the
discovery and deployment processes are executed in rounds. This is
sufficient since the goals of discovery and deployment are to find
installed servers, which does not change very frequently; hence it is
enough to run them in periodically occurring rounds.
As shown later in this Section, building of the network tree
requires collecting data from a possibly large number of
network nodes and domains, it is likely that the minimum duration of a round would be measured in minutes; we believe this to be sufficient for most use cases.

Structuring the whole process in rounds is actually beneficial since
it allows for all parties to determine the (temporal) extent of their
participation; the rounds define possible points at which to
change the service offerings. The first phase of the round creates a
map of the network with possible deployment points and the second
phase performs the actual deployment. Client redirection happens in
real-time and follows the setup of the current round.
Next, we examine the protocol flow in detail.

\subsection{Edge Server Discovery Phase}
\label{sec:edge-serv-disc}

The first phase of \edp aims to discover all possible edge servers associated on a client's path and build an aggregation tree composed of all clients. 
%
%
%
The phase is depicted in Figure~\ref{fig:sys-1-stage-1} and is summarized in Algorithm~\ref{alg:discovery}.
The explanation of steps follows below.

\begin{algorithm}[!tb]
  \caption{\edp discovery phase}
  \label{alg:discovery}
  \textbf{Step 1:} Collect client IP addresses from requests.
  
  \textbf{Step 2:} Run \texttt{traceroute}s to client IP addresses
  to build the router level tree. Identify domains of the routers.
  
  \textbf{Step 3:} Retrieve DNS SRV records for the identified
  domains to get addresses of possible edge servers in the tree.

  \textbf{Result:} A tree from the cloud to clients identifying
  locations of edge servers, their IP addresses, and port numbers.
\end{algorithm}

 
%
%

%
\noindent
\textbf{Step 1: Determine paths to clients}: For every client that
sends a request, the orchestrator sends out a \texttt{traceroute} to
identify the path from the cloud to the client. It must be noted, that \edp utilizes \texttt{traceroute} only as a tool to build a network graph of user to cloud connection which can be easily replaced by crowdsourced methodologies or active network sniffing tools such as \texttt{netstat}. We discuss the
limitations of the current approach in Section~\ref{sec:netw-topol-expl}; for
now, we assume \texttt{traceroute} to be sufficient for the functionality of \edp.

\noindent
\textbf{Step 2: Determine on-path DNS zones}: The \texttt{traceroute}s give us the routers on the path to the client and using either DNS PTR records or augmenting this with \texttt{whois} information for routers with no PTR records, we can identify the domains along the path.

\noindent
\textbf{Step 3: Locate the edge servers}: Once we have the list of domains, we perform a DNS SRV lookup using our new \texttt{edge} type SRV record defined below, which returns the addresses of edge servers in that domain, i.e., possible locations where services could be onloaded.

%
This step is central to the working of \edp discovery phase.
\edp relies on the assumption that majority of edge providers will add SRV \texttt{edge} type of record to DNS in attempt to gain from
the discovery of their edge infrastructure, but given that they wish to attract clients, they have a strong incentive to do so (and ensure the PTR records are accurate as well).
We provide examples of DNS SRV records, containing IP addresses of edge servers, in Figure~\ref{fig:srv-record}.
Each SRV record is arranged according to the following format:
\begin{lstlisting}
_service.  _protocol.name. TTL  DNSclass  SRV  priority  weight  port  target
\end{lstlisting}
Figure~\ref{fig:srv-record} shows that DNS zone \texttt{domainA} hosts two edge servers: \texttt{serverA} and \texttt{serverB}.
\begin{figure}[!tb]
\begin{lstlisting}
_edge._tcp.domainA.com. 86400 IN SRV 10 30 5060 serverA.domainA.com.
_edge._tcp.domainA.com. 86400 IN SRV 10 10 5060 serverB.domainA.com.

_edge._udp.domainA.com. 86400 IN SRV 10 30 1720 serverA.domainA.com.
_edge._udp.domainA.com. 86400 IN SRV 10 10 1720 serverB.domainA.com.

serverA.domainA.com.  86400 IN A 192.168.121.30
serverB.domainA.com.  86400 IN A 192.168.121.31
\end{lstlisting}
\caption{DNS SRV Record for Edge Servers}
\label{fig:srv-record}
\end{figure}
Servers support both TCP and UDP, but \texttt{serverB} should be
preferred (priority 10 vs. 30). With this information, the
orchestrator can build a graph (a tree) from its location to the
locations of all known clients, and indicate on the tree the possible
addresses of edge servers that would be on the path between the client and the cloud.
%
%
%


\subsection{Negotiation and Deployment Phase}
\label{sec:negot-depl-phase}

The second phase of \edp aims to onload the selected service on the edge servers located in the first phase. 
The communication flow is depicted in Figure~\ref{fig:sys-1-stage-2} and has been summarized in Algorithm~\ref{alg:deployment}.
The deployment phase of \edp is processed in three steps.

\begin{algorithm}[!tb]
  \caption{\edp deployment phase}
  \label{alg:deployment}
  \textbf{Step 1:} Identify which services should be onloaded to edge
  servers and onto which servers they should go.

  \textbf{Step 2:} Optionally negotiate deployment of services and
  revise placement if the original placement is not feasible (details not
  covered in this paper)

  \textbf{Step 3:} Deploy services, either containers or VMs. Redirect clients to deployed instances using HTTP 302 response.

  \textbf{Result:} Service containers or VMs deployed on selected edge servers, and clients are redirected correctly.
\end{algorithm}


\noindent
\textbf{Step 1: Identify candidate services for onloading}: Onloading starts by identifying the services whose relocation to the edge would be the most beneficial.
%
Figure~\ref{fig:sys-1-stage-2} shows cloud provider hosting two services, \texttt{X} and \texttt{Y}.
The services run, for example, as Docker containers, or inside
dedicated VMs; for both cases, there are efficient live migration
techniques \cite{docker_criu, Ismail2015EvaluationOD,
  Chaufournier2017}. 
%
The selection of service to onload can be based on multiple metrics, e.g., bandwidth consumed, processing requirements, QoS, geographical diversity of users subscribed to the service.
In Figure~\ref{fig:sys-1-stage-2}, the orchestrator determines to
onload service \texttt{X}.

%

%
\noindent
\textbf{Step 2: Server selection and service request}: Next, we select
candidate locations in the tree for deploying the services identified
in the first step. Considering the metrics used to determine the
services, we can similarly identify locations in the network tree for
deploying these services. This is a straight-forward (albeit possibly
complex) optimization process, specific to the particular service or
set of services; hence we omit the details of this process and simply
assume that it can be performed.

Next, we must ensure that the selected locations are available to host
the onloaded services. The cloud initiates a request to the selected
edge servers to ensure that they have sufficient capacity to host the
services. In case of insufficient capacity, we choose alternative locations.

\noindent
\textbf{Step 3: Deployment and redirection}: In this step, we migrate
the selected container to the decided edge server. 
For stateless services, we copy the required container or VM from cloud machine to the new host edge server.  
For live migration, various techniques have been developed for Docker containers~\cite{Ismail2015EvaluationOD, docker_criu, Ma2017EfficientSH} and VMs~\cite{Chaufournier2017, Ha2017}, considering specifically edge environments.

Clients are redirected using an appropriate HTTP response. When the
cloud provider wishes to redirect a client, instead of the normal
reply with the requested content, it replies with a \texttt{302 Moved
  temporarily} response URL that indicates the IP address and port of
the selected edge server. This can be augmented with a TTL-field to
ensure that the redirection is valid for only as long as the
deployment of the service on edge can be guaranteed (e.g., until
the end of this round of the protocol).

\section{\edp in Practice}
\label{sec:edp-practicalities}

As discussed in Section~\ref{sec:edp}, \edp relies on a graph composed
of network connections via intermediate edge servers from all clients
to the cloud. As the construction of the tree and its correctness is
paramount to the functioning of \edp, we now show how this tree can be constructed.

\begin{figure}[!tb]
	\centering
	\includegraphics[width=1\linewidth, height=0.8\linewidth]{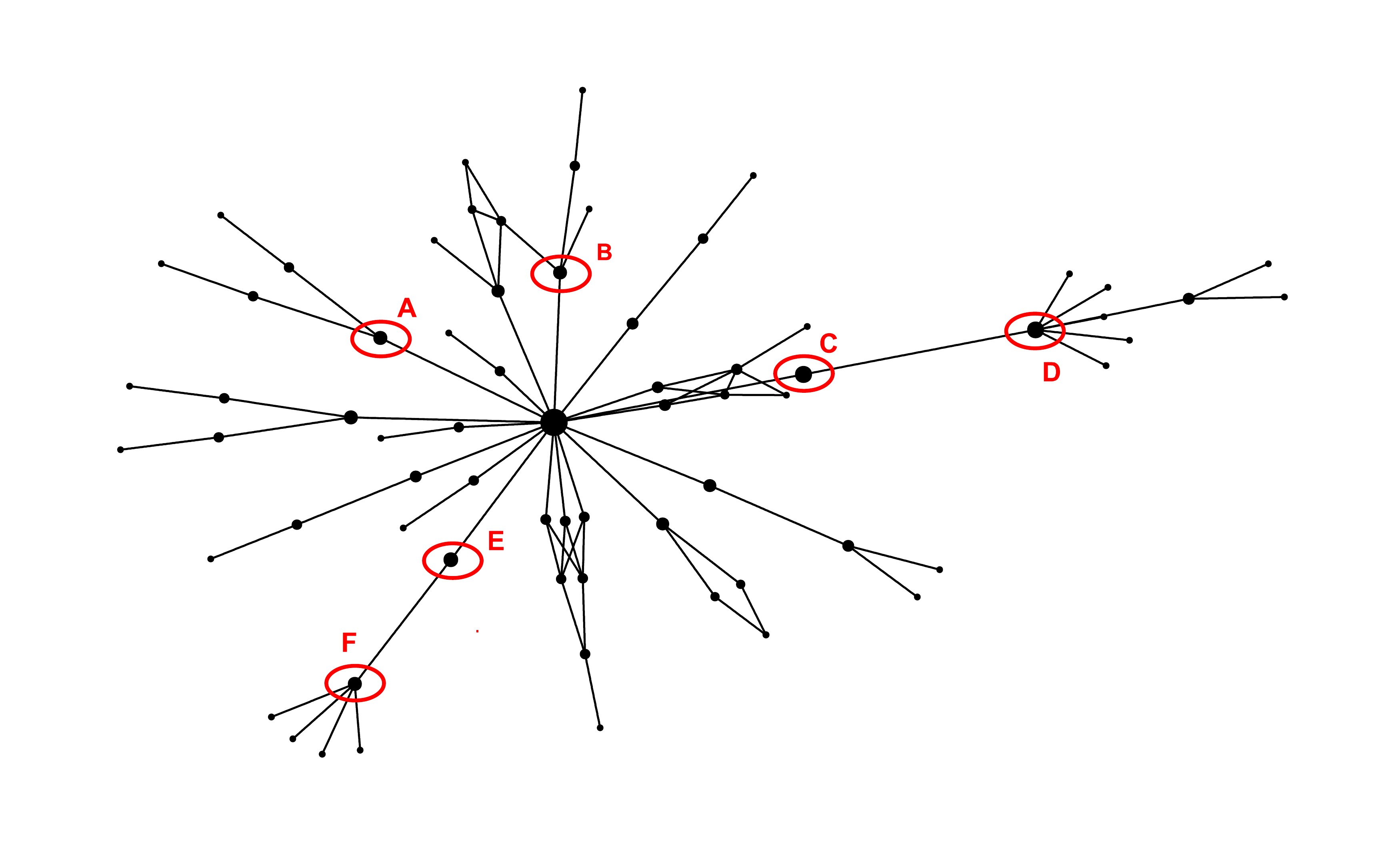}
	\caption{Network aggregate tree from cloud to clients via intermediate edge servers.}
	\label{fig:netw-graph}
\end{figure}

We ran our tests from an instance of Amazon Web Services (AWS) running
in Frankfurt, Germany. This node represents the cloud, and as clients
we picked the 100 universities from the Times Higher Education
ranking\cite{timesranking}. We chose the main websites of these
universities to represent the clients, to simulate a situation where the service running on the cloud has global interest. From the AWS instance, we ran \texttt{traceroute}s to these IP addresses and built the network graph. To simplify the graph, we grouped nodes into /24 subnets. Since there are no DNS SRV edge records, we merely assume every node in this graph would be a potential edge server location. Figure~\ref{fig:netw-graph} shows a subset of the tree.


The central node in the figure is the AWS instance from where all the
\texttt{traceroute}s were sent, and the other points are the discovered
routers or clients nodes. The size of a point indicates the
betweenness centrality of the node, i.e., gives an indication of how
many paths to clients pass through that node.\footnote{We only calculated the betweenness centrality for the paths originating at the center since those are the only paths of interest to us.}
Ideally, we should deploy edge servers on nodes with high betweenness
centrality (i.e., they serve many clients) and that are as close to
the clients as possible (i.e., to minimize latency). Balancing this
tradeoff is very much application specific, but it is something the
orchestrator would need to take into account when making onloading
decisions. For example, in the figure both nodes ``C'' and ``D'', as
well as ``E'' and ``F'' have the same betweenness centrality, but
nodes ``D'' and ``F'' should be preferred in deployment, as they are
closer to the clients at the edge.
 

\section{Discussion}
\label{sec:discussion}

We now discuss practical aspects of \edp and expand on its use cases.

\subsection{Network Topology Exploration}
\label{sec:netw-topol-expl}

\edp relies on \texttt{traceroute} for its exploration of the paths to the clients. It uses the ICMP protocol, and some routers choose to
ignore the ping messages sent by \texttt{traceroute}, resulting in us
getting an incomplete picture of the network. Likewise, not all
routers in the network have a DNS PTR record that would enable us to
map their IP addresses to domain names. While the \texttt{whois}
service helps in trying to map IP addresses to domains, it is not
guaranteed to provide a unique (and correct) domain name for a given
IP address.

However, we do not see this as a major issue for \edp. Our rationale
is that since the deployed edge servers are intended to be used, the
entity deploying them has a strong incentive to ensure that the DNS
records are correct and that the information is available. 

\subsection{Use Cases}
\label{sec:use-cases}

\edp is not limited to computational onloading that we have just
examined. As mentioned above, we chose that as the example scenario
because it showcases the full potential of \edp; other use cases we
discuss below are more straightforward to implement.

A typical scenario for edge computing is offloading, where a client
having limited capacity transfers computational tasks to an edge
server. In this case, the client simply queries its current domain for the DNS SRV edge records, getting a list of possible nearby edge servers as a result. While this solution does not require the client to use \texttt{traceroute}, it only enables discovery in the current
domain. As an extension, the client could explore the network via
\texttt{traceroute} towards a cloud in the network to get additional
domains, but this would induce additional overhead.

There are numerous showcases of contextual edge applications, such as providing 360\degree\xspace panoramic view at the stadium by aggregating individual users' video streams \cite{nokia_mec}, improving the customer experience at a shopping mall \cite{shopping_edge}, and others. 
The discovery of those applications is generally enabled by external
means, such as billboards, emails and so on, but with \edp, it is possible to query edge servers for contextual applications, in the manner described above.
%

\subsection{\edp Opportunities}
\label{sec:open}

Edge computing needs open and standardized infrastructures and
protocols in order for its potential to be truly realized. This
implies a world where clients and platform and service providers can
discover, deploy, and use services flexibly and at run-time.

We believe \edp is an excellent match for discovering edge services
since it requires no modifications to existing infrastructure or
clients. It works efficiently and enables providers to flexibly onload their services as well as client-side offloading and contextual edge service discovery. It is entirely based on existing protocols and can operate in any environment.

\edp provides the crucial technical building block for wide-spread
deployment of edge services, and it needs to be complemented with
additional work on determining agreements between the various
providers and addressing security issues that may arise in these kinds of environments.

\section{Related Work}
\label{sec:related}

Our primary use case, computational onloading, is discussed by~\cite{bhardwaj2016fast, esposito2017complete}, among others.
Bhardwaj et al.~\cite{bhardwaj2016fast} implement the edge discovery as a backend service that hosts a directory of available devices.
That requires a centralized catalog approach we choose to avoid.
In~\cite{esposito2017complete}, the discovery is seen as identification of available capacities from the set of already known resources.

Edge and middlebox discovery have strong commonalities. In fact, Detal et al.~\cite{detal2013} use modified {\tt traceroute} for identifying hidden middleboxes along a network path.
Abujoda et al.~\cite{abujoda2015midas} suggest a custom protocol for discovery of middleboxes, criticizing {\tt traceroute} for low speed.
As a drawback, all middleboxes need to run specific controller software to enable discovery.
Nevertheless, despite the issues with \texttt{traceroute}, we consider
the superior alternative since it obviates the need for installing
additional software.

Mobile Edge Computing (MEC)~\cite{etsi2014mobile, hu2015mobile, beck2014mobile, sabella2016mobile} is an ongoing standardization effort for edge computing in the area of mobile applications.
In MEC, there is a strong assumption that edge servers are located at
cellular base stations, which makes discovery easy, but their
operation may be difficult in the presence of end-to-end encrypted connections.
\edp allows for discovery in MEC as well, and it can be triggered by
the cloud.

Varghese et al. \cite{varghese2017edge} present a concept of an EaaS (Edge-as-a-Service) platform, offering a discovery protocol.
However, the approach relies on specific master controller nodes: to utilize edge, one must know the particular provider of EaaS beforehand to be able to discover the edge resources.
Work by A. Salem et al. \cite{salem2017kinaara} suggests \textit{Kinaara}, an edge computing framework also offering discovery capabilities.
Discovery is enabled by special mediator nodes, that keep a connection to the cloud.
%
They do not discuss how mediators are discovered.
%
Mortazavi et al. suggest the concept of \textit{path computing}~\cite{Mortazavi2017}, which is quite similar to our view of edge environment, where edge resources reside on a path from the cloud to an end user.
Their framework, \textit{CloudPath}, addresses many practical issues, but edge discovery does not happen on the fly: the developer is actually responsible for specifying the mappings between application functions and actual computing facilities in the \textit{deployment descriptor file}.
We suppose that frameworks presented in \cite{varghese2017edge}, \cite{Mortazavi2017}, and \cite{salem2017kinaara} can potentially gain from utilizing the \edp, which will remove the necessity for highly specialized components, adds more agility, and improves the overall user experience since \edp is capable of discovering previously unknown resources.
%

\section{Conclusion}
\label{sec:conclusion}

Discovery of edge services is a vital part of making edge computing a
reality. In this paper, we have presented \edp, a DNS-based edge
discovery protocol that enables efficient discovery of edge nodes,
both for service providers and clients. \edp uses existing protocols
and services and requires no modifications to the infrastructure or
client devices. We leverage standard tools, such as
\texttt{traceroute} and DNS and because of this, \edp can be deployed
in the networks as they are. We have shown a practical feasibility
study that demonstrates the basic concepts of \edp. As part of our
future work, we plan on implementing the service placement components
of \edp and make the solution available.



%

{\footnotesize \bibliographystyle{abbrv}
\bibliography{bib}}

\begin{thebibliography}{10}

\bibitem{abujoda2015midas}
A.~Abujoda and P.~Papadimitriou.
\newblock Midas: Middlebox discovery and selection for on-path flow processing.
\newblock In {\em Communication Systems and Networks (COMSNETS), 2015 7th
  International Conference on}, pages 1--8. IEEE, 2015.

\bibitem{Bahreini2017EfficientPO}
T.~Bahreini and D.~Grosu.
\newblock Efficient placement of multi-component applications in edge computing
  systems.
\newblock In {\em SEC}, 2017.

\bibitem{beck2014mobile}
M.~T. Beck, M.~Werner, S.~Feld, and S.~Schimper.
\newblock Mobile edge computing: A taxonomy.
\newblock In {\em Proc. of the Sixth International Conference on Advances in
  Future Internet}, pages 48--55. Citeseer, 2014.

\bibitem{bhardwaj2016fast}
K.~Bhardwaj, M.-W. Shih, P.~Agarwal, A.~Gavrilovska, T.~Kim, and K.~Schwan.
\newblock Fast, scalable and secure onloading of edge functions using airbox.
\newblock In {\em Edge Computing (SEC), IEEE/ACM Symposium on}, pages 14--27.
  IEEE, 2016.

\bibitem{bonomi2012fog}
F.~Bonomi, R.~Milito, J.~Zhu, and S.~Addepalli.
\newblock Fog computing and its role in the internet of things.
\newblock In {\em Proceedings of the first edition of the MCC workshop on
  Mobile cloud computing}, pages 13--16. ACM, 2012.

\bibitem{shopping_edge}
S.~Carlini.
\newblock {How the Internet of Things and Edge Computing Will Help
  Revolutionize the Shopping Experience}.
\newblock
  \url{https://blog.schneider-electric.com/datacenter/2016/06/03/iot-edge-computing/},
  2016.

\bibitem{Ceselli2017MobileEC}
A.~Ceselli, M.~Premoli, and S.~Secci.
\newblock Mobile edge cloud network design optimization.
\newblock {\em IEEE/ACM Transactions on Networking}, 25:1818--1831, 2017.

\bibitem{Chaufournier2017}
L.~Chaufournier, P.~Sharma, F.~Le, E.~Nahum, P.~Shenoy, and D.~Towsley.
\newblock Fast transparent virtual machine migration in distributed edge
  clouds.
\newblock In {\em Proceedings of the Second ACM/IEEE Symposium on Edge
  Computing}, SEC '17, pages 10:1--10:13, New York, NY, USA, 2017. ACM.

\bibitem{docker_criu}
{CRIU}.
\newblock {Docker - CRIU}.
\newblock \url{https://criu.org/Docker}, 2018.

\bibitem{detal2013}
G.~Detal, B.~Hesmans, O.~Bonaventure, Y.~Vanaubel, and B.~Donnet.
\newblock Revealing middlebox interference with tracebox.
\newblock In {\em Proceedings of the 2013 Conference on Internet Measurement
  Conference}, IMC '13, pages 1--8, New York, NY, USA, 2013. ACM.

\bibitem{esposito2017complete}
F.~Esposito, A.~Cvetkovski, T.~Dargahi, and J.~Pan.
\newblock Complete edge function onloading for effective backend-driven cyber
  foraging.
\newblock In {\em Wireless and Mobile Computing, Networking and Communications
  (WiMob),}, pages 1--8. IEEE, 2017.

\bibitem{etsi2014mobile}
ETSI.
\newblock {Mobile-Edge Computing}.
\newblock
  \url{https://portal.etsi.org/portals/0/tbpages/mec/docs/mobile-edge_computing_-_introductory_technical_white_paper_v1%2018-09-14.pdf},
  2014.

\bibitem{Ha2017}
K.~Ha, Y.~Abe, T.~Eiszler, Z.~Chen, W.~Hu, B.~Amos, R.~Upadhyaya, P.~Pillai,
  and M.~Satyanarayanan.
\newblock You can teach elephants to dance: Agile vm handoff for edge
  computing.
\newblock In {\em Proceedings of the Second ACM/IEEE Symposium on Edge
  Computing}, SEC '17, pages 12:1--12:14, New York, NY, USA, 2017. ACM.

\bibitem{hu2015mobile}
Y.~C. Hu, M.~Patel, D.~Sabella, N.~Sprecher, and V.~Young.
\newblock Mobile edge computing—a key technology towards 5g.
\newblock {\em ETSI white paper}, 11(11):1--16, 2015.

\bibitem{Ismail2015EvaluationOD}
B.~I. Ismail, E.~M. Goortani, M.~B.~A. Karim, W.~M. Tat, S.~Setapa, J.~Y. Luke,
  and O.~H. Hoe.
\newblock Evaluation of docker as edge computing platform.
\newblock {\em 2015 IEEE Conference on Open Systems (ICOS)}, pages 130--135,
  2015.

\bibitem{liu2017barriers}
T.~Liu, Z.~Tariq, J.~Chen, and B.~Raghavan.
\newblock The barriers to overthrowing internet feudalism.
\newblock In {\em Proceedings of the 16th ACM Workshop on Hot Topics in
  Networks}, pages 72--79. ACM, 2017.

\bibitem{Ma2017EfficientSH}
L.~Ma, S.~Yi, and Q.~Li.
\newblock Efficient service handoff across edge servers via docker container
  migration.
\newblock In {\em SEC}, 2017.

\bibitem{nokia_mec}
{Mobile Europe}.
\newblock {Nokia brings AR to sports stadium with MEC platform}.
\newblock
  \url{https://www.mobileeurope.co.uk/press-wire/nokia-bring-ar-to-sports-stadium-with-mec-platform},
  2017.

\bibitem{mohan2016edge}
N.~Mohan and J.~Kangasharju.
\newblock Edge-fog cloud: A distributed cloud for internet of things
  computations.
\newblock In {\em Cloudification of the Internet of Things (CIoT)}, pages 1--6.
  IEEE, 2016.

\bibitem{Mortazavi2017}
S.~H. Mortazavi, M.~Salehe, C.~S. Gomes, C.~Phillips, and E.~de~Lara.
\newblock Cloudpath: A multi-tier cloud computing framework.
\newblock In {\em Proceedings of the Second ACM/IEEE Symposium on Edge
  Computing}, SEC '17, pages 20:1--20:13, New York, NY, USA, 2017. ACM.

\bibitem{sabella2016mobile}
D.~Sabella, A.~Vaillant, P.~Kuure, U.~Rauschenbach, and F.~Giust.
\newblock Mobile-edge computing architecture: The role of mec in the internet
  of things.
\newblock {\em IEEE Consumer Electronics Magazine}, 5(4):84--91, 2016.

\bibitem{salem2017kinaara}
A.~Salem, T.~Salonidis, N.~Desai, and T.~Nadeem.
\newblock Kinaara: Distributed discovery and allocation of mobile edge
  resources.
\newblock In {\em Mobile Ad Hoc and Sensor Systems (MASS), 2017 IEEE 14th
  International Conference on}, pages 153--161. IEEE, 2017.

\bibitem{satyanarayanan2009case}
M.~Satyanarayanan, P.~Bahl, R.~Caceres, and N.~Davies.
\newblock The case for vm-based cloudlets in mobile computing.
\newblock {\em IEEE pervasive Computing}, 8(4), 2009.

\bibitem{timesranking}
{Times Higher Education}.
\newblock {World University Rankings}.
\newblock
  \url{https://www.timeshighereducation.com/world-university-rankings/2018/world-ranking},
  2018.

\bibitem{varghese2017edge}
B.~Varghese, N.~Wang, J.~Li, and D.~S. Nikolopoulos.
\newblock Edge-as-a-service: Towards distributed cloud architectures.
\newblock {\em arXiv preprint arXiv:1710.10090}, 2017.

\end{thebibliography}


\end{document}